\documentstyle[preprint,prc,aps,epsf]{revtex}

\draft

\newcommand{\vect}[1]{\mbox{\bf #1}}
\newcommand{\vecs}[1]{\mbox{\scriptsize \bf #1}}
\newcommand{\rmt}[1]{\mbox{\tiny #1}}

\begin{document}

%
\title{Analysis of correlation data of $\pi^+\pi^-$ pair in $p$ +
Ta reaction using Coulomb wave function
including momentum resolution and strong interaction effect}

\author{T. Mizoguchi$^1$\thanks{e-mail:
mizoguti@yukawa.kyoto-u.ac.jp},
M. Biyajima$^2$\thanks{e-mail: biyajima@azusa.shinshu-u.ac.jp},
I. V. Andreev$^3$\thanks{e-mail: andreev@lpi.ac.ru} and
G. Wilk$^4$\thanks{e-mail: wilk@fuw.edu.pl}}

\address{$^1$Toba National College of Maritime Technology, Toba
517-0012, Japan\\
$^2$Department of Physics, Faculty of Science, Shinshu
University, Matsumoto 390-0802, Japan\\
$^3$P. N. Lebedev Institute of Physics (FIAN), 117 924
Moscow, Russia\\
$^4$The Andrzej Soltan Institute for Nuclear Studies,
Nuclear Theory Department, Ho\.za 69, PL-00-681 Warsaw, Poland}

\date{\today}

\maketitle

%
\begin{abstract}
We have proposed a new method for the Coulomb wave function
correction which includes the momentum resolution for charged hadron
pairs and applied it to the precise data on $\pi^+\pi^-$ correlations
obtained in $p + \mbox{Ta}$ reaction at 70 GeV/c. It is found that
interaction regions of this reaction (assuming Gaussian source
function) are $9.8\pm 5.8$ fm and $7.7\pm 4.8$ fm for the thicknesses
of the target 8 $\mu$m and 1.4 $\mu$m, respectively. We have also
analyzed the data by the formula including strong interaction effect.
The physical picture of the source size obtained in this way is
discussed.
\end{abstract}


\narrowtext

%
\section{Introduction}
We have obtained recently new formulae for the Coulomb wave function
corrections for charged hadron pairs\ \cite{biya95,biya96}. Among
other things we have applied them \cite{biya96} to data on
$\pi^+\pi^-$ correlation obtained in $p + \mbox{Ta}$ reaction at 70
GeV/c (and corrected by the usual Gamow factor only). However, as was
pointed out to us by one of the author \cite{afan95}, our approach
does not account for the finite momentum resolution published in
\cite{afan90}. In fact, our formulae cannot be applied directly to
experimental data in which such momentum resolution is accounted for.
Therefore in the present work we would like to extend our method for
the Coulomb wave correction provided in \cite{biya95,biya96} in such
way as to include also the momentum resolution case and we would like
to re-analyse data of \cite{afan90} as well as the new data of
\cite{afan96} obtained with two kinds of thickness of the Ta target:
8 $\mu$m and 1.4 $\mu$m.

In the next Section we provide, for the sake of completeness, the
main
points of the analysis method used in Ref.\ \cite{afan90} whereas in
the Section 3 we outline our new method for the Coulomb wave
correction including this time also the momentum resolution. In
Section 4 we consider also the effect of the strong final state
interactions. The final part contains our concluding remarks.

%
\section{Data reconstruction method (using the Gamow factor with
momentum resolution)}
\begin{figure}\epsfxsize=16cm
  \centerline{\epsfbox{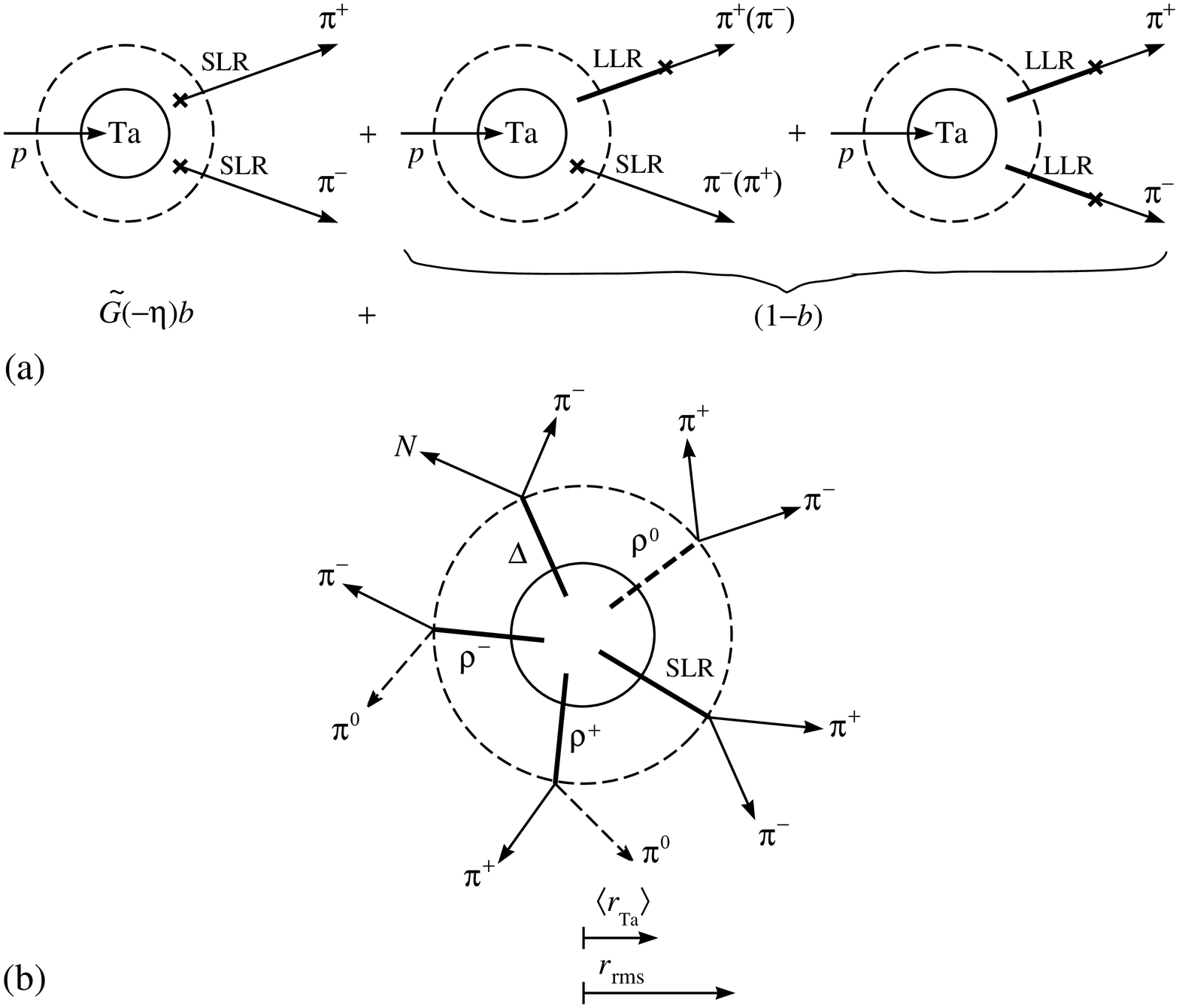}}
  \caption{$(a)$ Physical picture of Eq.\ (\protect\ref{eqn:a1}).
  $(b)$ Enlarged physical picture of interaction region.}
  \label{fig1}
\end{figure}

It has been stressed in\ \cite{afan90} that relative momenta of
$\pi^+\pi^-$ pairs observed by them have some finite resolutions. The
averaged correlation function, defined as
\begin{equation}
  R(\pi^+\pi^-) = \sigma \frac{d^2\sigma}{dp_1dp_2} \left/ \left(
  \frac{d\sigma}{dp_1} \frac{d\sigma}{dp_2}\right)\right.
  = G_{\rmt C} b + (1-b)\Phi (q),
  \label{eqn:a1}
\end{equation}
depends therefore on this momentum resolution, where $G_{\rmt C}$
denotes a correlation function of $\pi^+\pi^-$ pairs from decays of
short-lived resonances\ (SLR), and $\Phi(q)$ denotes a correlation
function between pions from decays of the long-lived resonances\
(LLR)
or LLR's and SLR's. The parameter $b$ means the portion of the
correlation function governed by the Coulomb wave function. For the
sake of simplicity $\Phi(q) = 1$. Fig.\ \ref{fig1}(a) shows physical
picture of Eq.\ (\ref{eqn:a1}). To account for it the following
random number method has been proposed in Ref.\ \cite{afan90} in
order
to obtain the corresponding averaged quantities in analysis of the
correlation data:
\begin{itemize}
\item[$(i)$] The relative momentum of the measured pair, $q = p_1 -
p_2$, is decomposed into its longitudinal and transverse components:
$q_{\rmt L}$ and $q_{\rmt T}$, respectively, by making use of the
uniform random number $u\in (0,1)$ (it is worthwhile to notice at
this point that the transverse components $q_{\rmt T}$ in data of
\cite{afan90} are smaller than $10$ MeV/c). The following scheme has
been employed:
\begin{mathletters}
  \label{eq:a2}
\begin{eqnarray}
  & & q_{\rmt T} =
  \left\{\begin{array}{l}
    10\sqrt u \quad\mbox{for}\quad q \ge 10\mbox{MeV/c}\; ,\\
    q\sqrt u \quad\mbox{for}\quad q \le 10\mbox{MeV/c}\; ,
  \end{array}\right.
  \label{eq:a2-a}\\
  & & q_{\rmt L} = \sqrt{q^2 - q_{\rmt T}^2}\:.
  \label{eq:a2-b}
\end{eqnarray}
\end{mathletters}
\item[$(ii)$] The Gaussian random numbers for $q_{\rmt L}$ and
$q_{\rmt T}$ are generated in the following way:
\begin{eqnarray}
  q_{\rmt L{\rm (random)}} &=& \sigma_{\rmt L} X + q_{\rmt L}\:,
  \label{eq:a3}\\
  q_{\rmt T{\rm (random)}} &=& \sigma_{\rmt T} X + q_{\rmt T} .
  \label{eq:a4}
\end{eqnarray}
where $X$ stands for the standard Gaussian random number \cite{FOOT1}
whereas $\sigma_{\rmt L}$ and $\sigma_{\rmt T}$ are longitudinal and
transverse setup resolutions for the corresponding components, which
are equal to (values used in\ \cite{afan96}): $\sigma_{\rmt L} = 1.3$
MeV/c; $\sigma_{\rmt T} = 0.6$ MeV/c (for target of the thickness 8
$\mu$m ) and $\sigma_{\rmt T} = 0.4$ MeV/c (for the 1.4 $\mu$m
target).
\item[$(iii)$] Using the randomized number $q_{\rm random} =
\sqrt{q_{\rmt L{\rm (random)}}^2 + q_{\rmt T{\rm (random)}}^2}$ one
calculates the corresponding randomized Gamow factor correction:
\begin{equation}
  G(-\eta_{\rm random}) =
  \frac{-2\pi\eta_{\rm random}}{\exp(-2\pi\eta_{\rm random})-1}\:,
  \label{eqn:a5}
\end{equation}
where $\eta_{\rm random} = m\alpha/q_{\rm random}$.
\end{itemize}
Calculating now the average value of $G(-\eta_{\rm random})$ in
$10\sim 100$ k events one can estimate the Gamow factor with this
finite momentum resolution,
\begin{equation}
  R(q) = \widetilde{G}(-\eta)b + (1-b)\:,
  \label{eqn:a6}
\end{equation}
where $b$ is the portion from the SLR's. It is understood (or,
rather, implicitly assumed) that the second term of light hand side,
$(1-b)$ originates from decays of LLR's like $\eta$, $K_0^S$,
$\Lambda$, and so on \cite{FOOT2}. Table\ \ref{tbl:a1} shows the
results of analysis of new data (for 8 $\mu$m target)\ \cite{afan96}
for region $q > 3$ MeV/c using this method\ \cite{FOOT3}.

%
\section{New method of Coulomb wave function corrections}
\begin{figure}\epsfysize=20cm
  \centerline{\epsfbox{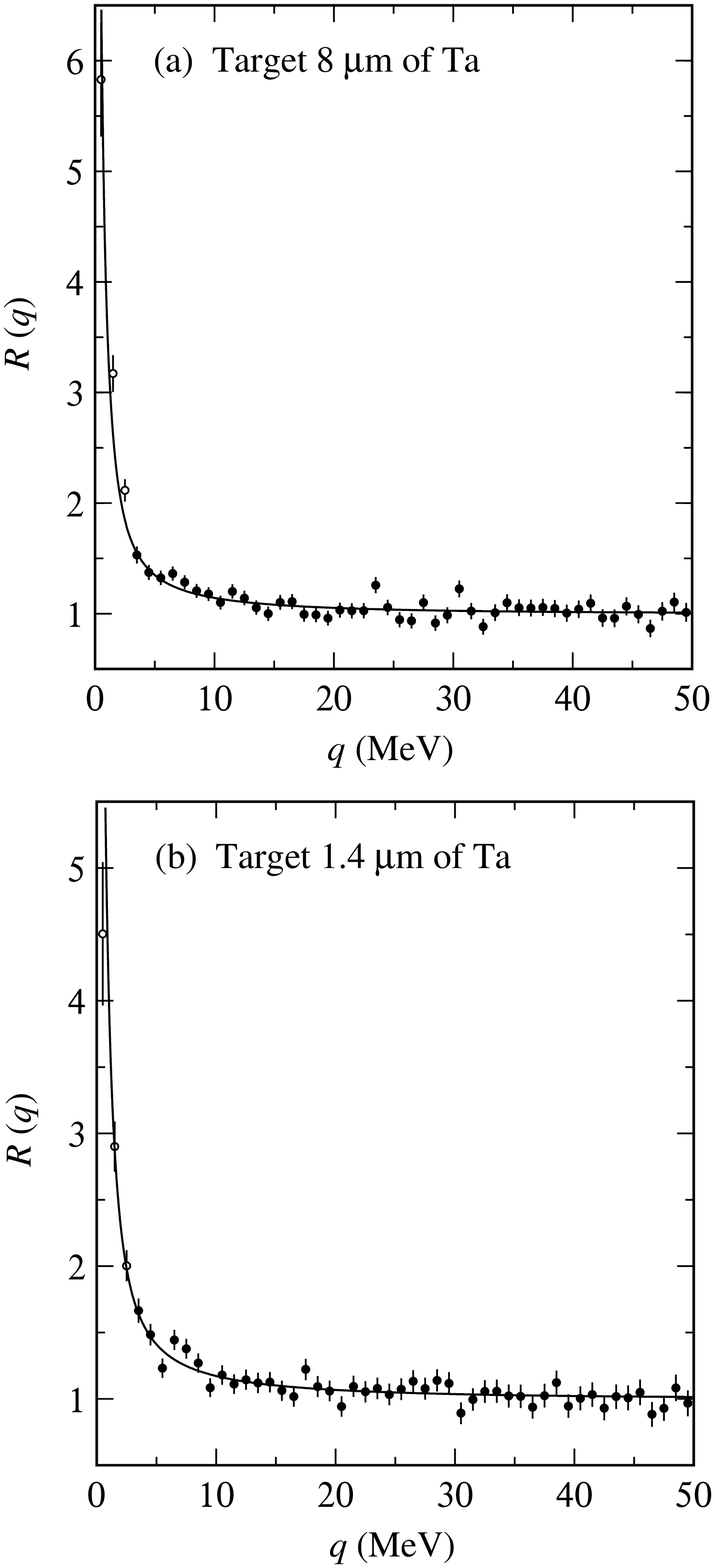}}
  \caption{Results of the $\chi^2$ fits for $p$ + Ta $\to
  \pi^+\pi^- + X$ reactions with $q > 3$ MeV/c by
  Eq.\ (\protect\ref{eqn:a11}) with (\protect\ref{eqn:a7}):
  $(a)$ 8 $\mu$m target; $(b)$ 1.4 $\mu$m target.}
  \label{fig2}
\end{figure}

We would like to propose the following new method of Coulomb wave
function correction (with a source function $\rho(r)$) to be used
instead of the Gamow factor and apply it to the same data as above.
As usual we decompose the wave function of unlike charged bosons with
momenta $p_1$ and $p_2$ into the wave function of the center-of-mass
system (c.m.) with total momentum $P = \frac{1}{2}(p_1 + p_2)$ and
the inner wave function with relative momentum $q = (p_1 - p_2)$.
This allows us to express Coulomb wave function
$\Psi_{\rmt C}(\vect{q},\;\vect{r})$ in terms of the confluent
hypergeometric function $F$\ \cite{schiff}:
\begin{equation}
 \Psi_{\rmt C}(\vect{q},\vect{r}) = \Gamma(1-i\eta)e^{\pi \eta/2}
e^{i\vecs{q}\cdot\vecs{r}/2} F(i\eta;1;iqr(1 - \cos
\theta)/2)\:,\\
  \label{eqn:a7}
\end{equation}
where $r = r_1 - r_2$ and the parameter $\eta = m\alpha/q$. Assuming
factorization in the source functions, $\rho(r_1)\rho(r_2) =
\rho_{\rmt R}(R)\rho_{\rm r}(r)$ (with $R = \frac{1}{2}(r_1 + r_2)$,
$\int \rho_{\rmt R}(R) d^3R = 1$ is assumed), we obtain the
expression
for Coulomb correction for the system of $\pi^+\pi^-$ pairs identical
(modulo the sign) as in Refs.\ \cite{bowler91,biya95,biya96}:
\begin{eqnarray}
  C_{\rmt C}(-\eta) &=& \int \rho_{\rmt R}(R) d^3R
  \int \rho_{\rm r}(r) d^3r|\Psi_{\rmt C}(\vect{q},\vect{r})|^2\\
  \label{eqn:a8-1}
  &=& G(-\eta)\sum_{n=0}^{\infty} \sum_{m=0}^{\infty}
  \frac{(-i)^n(i)^m}{n+m+1}\, q^{n+m}\, I(n,m) A_n A_m^*
  \nonumber\\
  &=& G(-\eta)[1 + \Delta_{1\rmt C}(-\eta)]\:,
  \label{eqn:a8-2}
\end{eqnarray}
where
$$
I(n,m) = 4 \pi \int dr\, r^{2 + n + m} \rho (r),\quad A_n =
\frac{\Gamma(-i\eta + n)}{\Gamma(-i\eta)}\frac{1}{(n!)^2}\:.
$$
For the specific choice of Gaussian source distribution, $\rho_r(r) =
\frac{\beta^3}{\sqrt{\pi^3}}\exp(-\beta^2 r^2)$, we have
\begin{eqnarray}
  I^{\rmt G} (n,m) &=& \frac 2{\sqrt{\pi}}
  \left(\frac 1{\beta}\right)^{n+m}
\Gamma\left(\frac{n+m+3}2\right),\\
  \label{eqn:a9}
  \Delta_{\rmt{1C}}(-\eta) &=& \frac{4\eta}{\sqrt{\pi}}
  \sum_{n=0}^{\infty} \frac{(-1)^n(n+1)!}{(2n+2)!(2n+1)}
  \left(\frac{q}{\beta}\right)^{2n+1}\:.
  \label{eqn:a9-2}
\end{eqnarray}
Using now the same method of Gaussian random numbers as in a
previous Section, we can analyse the old and the new data on
$\pi^+\pi^-$ pairs\ \cite{afan90,afan96} using the following formula:
\begin{equation}
  R(q) = \widetilde{C}_{\rmt C}(-\eta)b + (1-b)\:.
  \label{eqn:a11}
\end{equation}
Fig.\ \ref{fig2} and Table\ \ref{tbl:a1} show our results obtained
using Eq.\ (\ref{eqn:a8-2}) applied to old and new data with $q > 3$
MeV/c. Notice that now we can estimate (albeit with large errors)
also the dimension of the reaction region, not available when using
only Gamow factor.

%
\section{Strong interaction effect}
We shall consider now the
possible effect of strong final state interactions, using a
formulation proposed by Bowler in Ref.\ \cite{bowler88}. An extended
formula for charged particles has been proposed in Ref.\
\cite{osa96}.
As far as our knowledge is correct, data of phase shift of
$p$-wave {\sl at very small} $q$ region have not reported. Thus we
consider only $s$-wave. The wave function including Coulomb effect and
strong interaction effect of $s$-wave is described by
\begin{equation}
  \Psi_{\rm total}(\vect{q},\vect{r})=
  \Psi_{\rmt C}(\vect{q},\vect{r})
  + \Phi_{\rm st}(\vect{q},\vect{r})\ .
  \label{eqn:b1}
\end{equation}
Here $\Phi_{\rm st}(\vect{q},\vect{r})$ stands for the wave
function induced by the strong interactions\ \cite{osa96} and is
given by
\begin{mathletters}
  \label{eqn:b2}
\begin{equation}
  \Phi_{st}(\vect{q},\vect{r}) =
  \frac{\phi_0^{(2,0)}(\theta)[G_0(-\eta;\;qr/2) +
iF_0(-\eta;\;qr/2)]}{r}
  \label{eqn:b2-a}
\end{equation}
\begin{equation}
  \stackrel{\mbox{\scriptsize large } r}{\longrightarrow}
  \frac{\phi_0^{(2,0)}(\theta)\exp[\ i(qr/2+\eta\ln(qr) +
  \sigma_0)\ ]}{r},
  \label{eqn:b2-b}
\end{equation}
\end{mathletters}
where $\sigma_0 = \mbox{arg}\Gamma(1-i\eta)$, $F_0$ and $G_0$ are the 
regular and irregular solutions ($l=0$; $s$-wave) of the radial 
coordinates including the Coulomb potential\ \cite{FOOT4}. 
Eq.\ (\ref{eqn:b2}) is the scattering amplitude and the asymptotic 
form\ \footnote{Proof of Eq.\ (\ref{eqn:b1}) with Eq.\ (\ref{eqn:b2}) 
is given as follows:
$$
  e^{i\sigma_0}e^{i\delta_0}(\cos\delta_0\cdot F_0 
  + \sin\delta_0\cdot G_0)
  = e^{i\sigma_0}[F_0 + e^{i\delta_0}\sin\delta_0
  \cdot (G_0 + iF_0)].
$$
Using the above equality, and summing up all the partial waves with 
$\sigma_l = \mbox{arg}\Gamma(l+1-i\eta)$, we have Eq.\ (\ref{eqn:b1});
\begin{eqnarray*}
  &&[e^{i\sigma_0}F_0 + \sum_{l=1}^{\infty} i^l(2l+1) e^{i\sigma_l} 
  F_l(-\eta;\;qr/2) P_l(\cos\theta)\\
  && + e^{i\sigma_0}e^{i\delta_0} 
  \sin\delta_0\cdot (G_0 + iF_0)]/(qr/2)\\
  && = \Psi_{\rmt C}(\vect{q},\vect{r})
  + \Phi_{\rm st}(\vect{q},\vect{r})\ .
\end{eqnarray*}
}. We 
assume that an interpolation to the internal region
$[0,\ r_c \approx (1-2) \mbox{ fm}]$ is possible as in
Refs.\ \cite{bowler88,osa96}. (In other words, we assume a small
contribution from the internal region in the spatial integration. In
the present calculation, we do not introduce the cutoff factor
$r_c$.)

\begin{figure}\epsfysize=20cm
  \centerline{\epsfbox{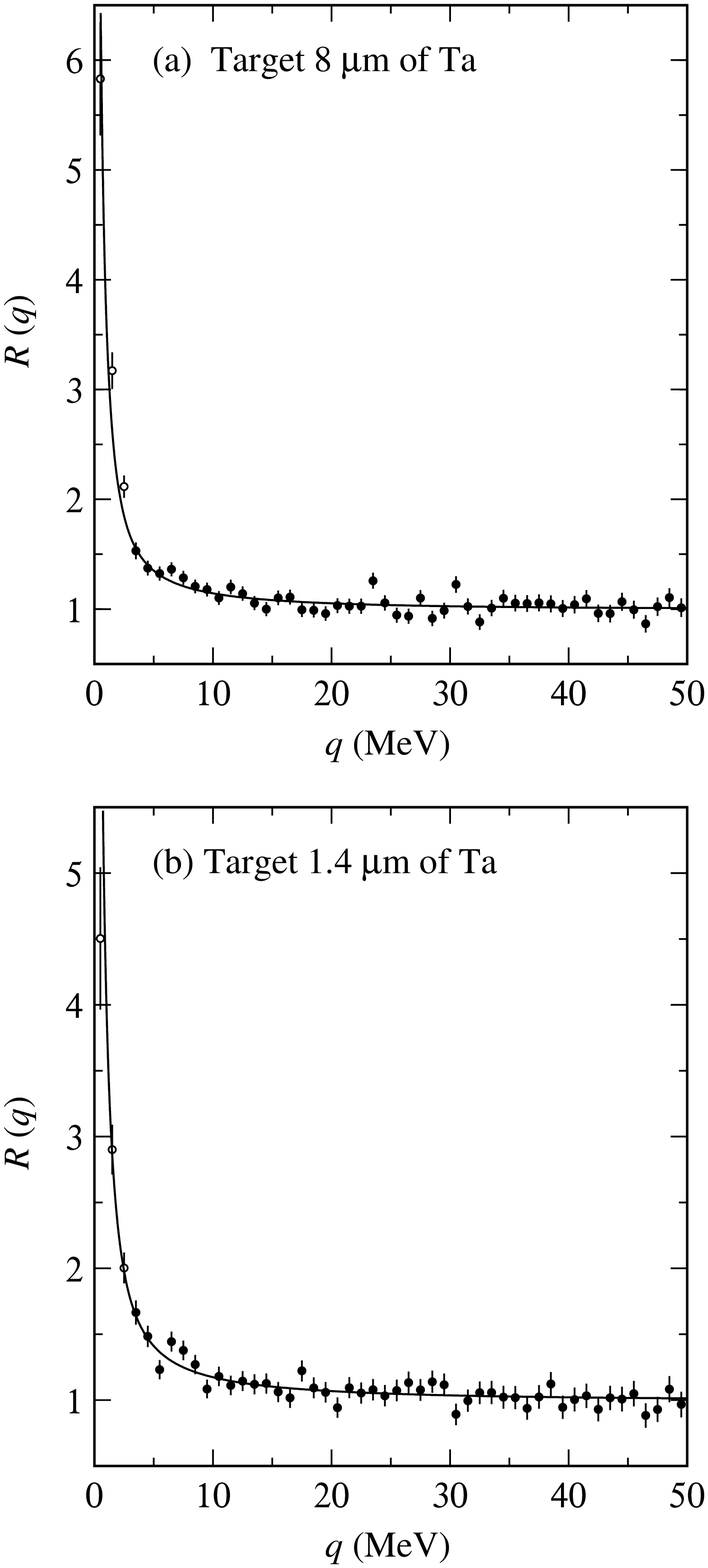}}
  \caption{Results of the $\chi^2$ fits for $p$ + Ta
  $\to \pi^+\pi^- + X$ reactions with $q > 3$ MeV/c by
  Eq.\ (\protect\ref{eqn:a11}) with (\protect\ref{eqn:b1}):
  $(a)$ 8 $\mu$m target; $(b)$ 1.4 $\mu$m target.}
  \label{fig3}
\end{figure}

The amplitude $\phi_0^{(2,0)}$ is decomposed as
\begin{mathletters}
  \label{eqn:b3}
\begin{eqnarray}
  && \phi_0^{(2,0)}  = \frac{1}{3} f_0^{(2)}(\theta)
  + \frac{2}{3}f_0^{(0)}(\theta)\ ,
  \label{eqn:b3-a}\\
  && f_0^{(2)}(\theta) =\frac{1}{iq}
  e^{i\sigma_0}[\ \exp(2i\delta^{(2)}_0)-1\ ]\ ,
  \label{eqn:b3-b}\\
  && f_0^{(0)}(\theta) =\frac{1}{iq}
  e^{i\sigma_0}[\ \exp(2i\delta^{(0)}_0)-1\ ]\ .
  \label{eqn:b3-c}
\end{eqnarray}
\end{mathletters}
For the phase shift of $I = 2$ and $0$ channel $\pi\pi$ scattering,
we use the following parameterization\ \cite{suzuki87}\ :
\begin{mathletters}
  \label{eqn:b4}
\begin{equation}
  \delta_0^{(2)}(q) = \frac{1}{2}\frac{a_0^{(2)}q}{1 +
0.5\mbox{[GeV$^{-2}$]}\,q^2},
  \quad a_0^{(2)} = -1.20\mbox{ GeV$^{-1}$},
  \label{eqn:b4-a}
\end{equation}
\begin{equation}
  \delta_0^{(0)}(q) = \frac 12 a_0^{(0)} q,
  \quad a_0^{(0)} = 1.50\mbox{ GeV$^{-1}$}.
  \label{eqn:b4-b}
\end{equation}
\end{mathletters}
Figs.\ \ref{fig3} show results of our analysis of the new data. In
table\ \ref{tbl:a1}, we also show our results obtained using
Eq.\ (\ref{eqn:b1}) instead of Eq.\ (\ref{eqn:a7}) in
Eq.\ (\ref{eqn:a8-1}). As can be seen we observe systematically
larger
values of the interaction region with strong final state interactions
included.

%
\section{Concluding remarks}
\begin{figure}\epsfysize=16cm
  \centerline{\epsfbox{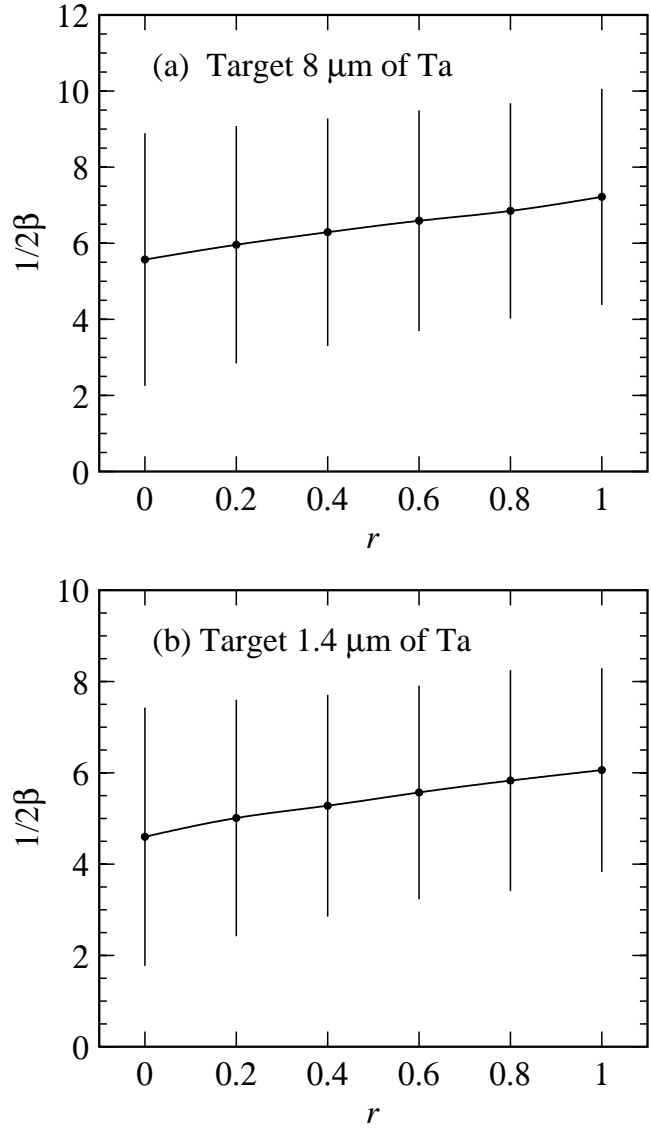}}
  \caption{Phase shift dependence of $1/2\beta$. The horizontal axis
  expresses strength of $a_0^{(2)}$ and $a_0^{(0)}$: $(a)$ 8 $\mu$m
  target; $(b)$ 1.4 $\mu$m target, $r a_0^{(2)}$ and $r a_0^{(0)}$
  $(0\le r \le 1.0)$.}
  \label{fig4}
\end{figure}

We have proposed new method for the Coulomb wave function correction
with momentum resolution and applied it to the analysis of the
precise data provided by\ Refs.\ \cite{afan90,afan96}. Authors of
Ref.\ \cite{afan90} have analysed their $\pi^+\pi^-$ correlation data
using Gamow factor for Coulomb corrections together with the random
numbers method to account for final momentum resolution. We have
repeated this analysis replacing Gamow factor by the Coulomb wave
function but following the same method for correction for the
momentum resolution effect (cf. Eq.\ (\ref{eqn:a8-2})). As a result
we were able to estimate the range of interaction for which (more
precisely, for the root mean squared ranges of interaction
$r_{\rm rms}$) we have obtained the following values (for the
Gaussian source function):
\begin{eqnarray}
  r_{\rm rms} = \frac{\sqrt 3}{2\beta} =
  \left\{\begin{array}{l}
    9.8\pm 5.8 \mbox{fm\quad for 8 $\mu$m}\; ,\\
    7.7\pm 4.8 \mbox{ fm\quad for 1.4 $\mu$m}\; .
  \end{array}\right.
  \label{eqn:a12}
\end{eqnarray}
{}From formula including also strong final state interaction effect
(cf. Eq.\ (\ref{eqn:b1})) we have obtained
\begin{eqnarray}
  r_{\rm rms} = \frac{\sqrt 3}{2\beta} =
  \left\{\begin{array}{l}
    12.5\pm 4.9 \mbox{fm\quad for 8 $\mu$m},\\
    10.5\pm 3.9 \mbox{ fm\quad for 1.4 $\mu$m}.
  \end{array}\right.
  \label{eqn:a13}
\end{eqnarray}
Values of Eq.\ (\ref{eqn:a13}) are fairly bigger than those of
Eq.\ (\ref{eqn:a12}). This fact means that $r_{\rm rms}$ estimated
by means of Eq.\ (\ref{eqn:b1}) depends strongly on phase shift
values. The dependence is shown in Figs.\ \ref{fig4}. The results are
also influenced by method of momentum resolution. To confirm
Eq.\ (\ref{eqn:a13}), we need to investigate different approaches in
a
future, for example, Ref.\ \cite{pratt90}. Table\ \ref{tbl:a2} shows
the results concerning cutoff dependences of momentum resolution.

The present study of $\pi^+\pi^-$ pair correlations has shown
therefore that one can estimate the interaction region even from the
$\pi^+\pi^-$ correlation data (although present data lead to large
errors for $r_{\rm rms}$). It can be compared with the size of
the Ta nucleus, which is given by:
\begin{eqnarray}
  \langle r_{\rmt{Ta}}\rangle &=& 1.2\times A^{1/3}
  \nonumber\\
   &=& 1.2\times (181)^{1/3} = 6.8 \mbox{ fm}\ .
  \label{eqn:a14}
\end{eqnarray}
As one can see, $r_{\rm rms}$ is significantly bigger than
$\langle r_{\rmt{Ta}}\rangle$. We attribute this difference to a
physical picture shown in Fig.\ \ref{fig1}(b), i.e., to the fact that
unlike-sign pions are about 60 \% emerging from the SLR's
($\varphi$, $\Delta$, $\cdots$) shown there. In a future one should
apply our theoretical formula to other data and estimate $\beta$'s
and $b$'s in similar reactions, and also consider a possibility of
more direct estimation of the parameter $b$ and its role in
determining the source size parameter\ \cite{FOOT5}.

%
\section*{Acknowledgements}
Authors are grateful to Dr. L. G. Afanas'ev for his kind
correspondences and for providing us with the new data on
$\pi^+\pi^-$ correlation prior to publication. They also sincerely 
thank a referee for its suggestion concerning Eq. (\ref{eqn:b2}).
This work is partially supported by the Grant-in-Aid for Scientific
Research from the Ministry of Education, Science, Sports and Culture
of Japan (\# 09440103) and (\# 08304024), the Japan Society of
Promotion of Science (JSPS), and the Yamada Foundation. One of authors 
(I. A.) is also partially supported by Russian Fund of Fundamental 
Research (grant 96-02-16347a). Numerical computations in this work 
are partially carried out at RCNP(Research Center for Nuclear Physics, 
Osaka Univ.) Computer Facility.

%

%

\newpage

\widetext

%
\begin{table}
\caption[table1]{Results of the $\chi^2$ fits of $R(q)$ for Gaussian
source by Eqs.\ (\ref{eqn:a6}) and (\ref{eqn:a11}).}
\label{tbl:a1}
\begin{tabular}{ccccc}
Reaction & Formula
& $1/2\beta$ [fm] & $b$ & $\chi^2/NDF$\\
\hline
data of Ref.\ \cite{afan90} (cf. Ref.\ \cite{biya96}) & Gamow factor
& --- & --- & 57.8/40\\
{\footnotesize (without momentum resolution)} & Eq.\ (\ref{eqn:a8-2})
& $2.30\pm 0.88$ & --- & 51.0/39\\
\hline
data of Ref.\ \cite{afan90} & Eq.\ (\ref{eqn:a6})
& --- & $1.0$ (fixed) & 53.6/37\\
{\footnotesize (with momentum resolution)}
& Eq.\ (\ref{eqn:a11}) with (\ref{eqn:a7}) &
$2.93\pm 1.03$ & $1.0$ (fixed) & 44.7/36\\
& Eq.\ (\ref{eqn:a11}) with (\ref{eqn:b1})
& $4.63\pm 0.75$ & $1.0$ (fixed) & 44.8/36\\
\hline\hline
data of Ref.\ \cite{afan96}, 8 $\mu$m & Eq.\ (\ref{eqn:a6})
& --- & $0.43\pm 0.03$ & 55.0/46\\
& Eq.\ (\ref{eqn:a11}) with (\ref{eqn:a7})
& $5.63\pm 3.34$ & $0.53\pm 0.07$ & 51.5/45\\
& Eq.\ (\ref{eqn:a11}) with (\ref{eqn:b1})
& $7.22\pm 2.84$ & $0.54\pm 0.07$ & 51.3/45\\
\hline
data of Ref.\ \cite{afan96}, 1.4 $\mu$m & Eq.\ (\ref{eqn:a6})
& --- & $0.51\pm 0.04$ & 37.9/46\\
& Eq.\ (\ref{eqn:a11}) with (\ref{eqn:a7})
& $4.44\pm 2.80$ & $0.60\pm 0.07$ & 35.1/45\\
& Eq.\ (\ref{eqn:a11}) with (\ref{eqn:b1})
& $6.06\pm 2.23$ & $0.61\pm 0.07$ & 35.1/45\\
\end{tabular}
\end{table}

%
\begin{table}
\caption[table1]{Momentum resolution and cutoff.}
\label{tbl:a2}
\begin{tabular}{ccccc}
Data     & using Monte Carlo events & $1/2\beta$
           [fm] & $b$ & $\chi^2/NDF$\\
\hline
data of Ref.\ \cite{afan96}, 8 $\mu$m & without momentum resolution &
$7.95\pm 3.05$ & $0.58\pm 0.08$ & 50.5/45\\
& (central value is used)\\
& $|q_{\rmt{L(T)}{\rm (random)}}-q_{\rmt{L(T)}}| \le
\sigma_{\rmt{L(T)}}$
& $7.66\pm 2.96$ & $0.56\pm 0.08$ & 50.7/45\\
& $|q_{\rmt{L(T)}{\rm (random)}}-q_{\rmt{L(T)}}| \le
2\sigma_{\rmt{L(T)}}$
& $7.48\pm 2.91$ & $0.55\pm 0.08$ & 51.1/45\\
& all & $7.22\pm 2.84$ & $0.54\pm 0.07$ & 51.3/45\\
\hline
data of Ref.\ \cite{afan96}, 1.4 $\mu$m & without momentum resolution
& $6.45\pm 2.36$ & $0.64\pm 0.08$ & 34.7/45\\
& (central value is used)\\
& $|q_{\rmt{L(T)}{\rm (random)}}-q_{\rmt{L(T)}}| \le
\sigma_{\rmt{L(T)}}$
& $6.29\pm 2.30$ & $0.63\pm 0.08$ & 34.6/45\\
& $|q_{\rmt{L(T)}{\rm (random)}}-q_{\rmt{L(T)}}| \le
2\sigma_{\rmt{L(T)}}$
& $6.18\pm 2.27$ & $0.62\pm 0.08$ & 35.3/45\\
& all & $6.06\pm 2.23$ & $0.61\pm 0.07$ & 35.1/45\\
\end{tabular}
\end{table}

\narrowtext


\begin{references}
\bibitem{FOOT1} The probability density of such random numbers
                  is given by
                  $$
                  f(x) = \frac 1{\sqrt{2\pi}} \exp(-x^2/2)
                  \quad (-\infty < x < \infty)\:.
                  $$
\bibitem{FOOT2} This fact has some consequences on the
                  determination of the radius of the interaction
                  region as will be shown later on and makes it
                  different from that obtained from the Bose-Einstein
                  correlations.
\bibitem{FOOT3} In the region $q < 3$ MeV/c Coulomb enhancement is
                  influenced by Coulomb bound state of $\pi^+\pi^-$
                  pairs (see Ref.\ \cite{afan96}). Then we use data
                  with $q > 3$ MeV/c in our analyses.
\bibitem{FOOT4} Authors of Ref.\ \cite{osa96} used the form of
                  Eq.\ (\ref{eqn:b2-b}) with $\sqrt{G(-\eta)}$
                  factor.
                  However, Eq. (\ref{eqn:b2-a}) is more precise in
                  small $q$ region than Eq.\ (\ref{eqn:b2-b}).
\bibitem{FOOT5} In the analysis of $\pi^0\pi^0$ correlation data,
                  a similar function $f(q)$ is introduced:
                  $$
                  R(\pi^0\pi^0) = f(q) +
                  (1 - f(q))[1 + \lambda E_{2\rmt B}^2]\:,
                  $$
                  where $\lambda$ and $E_{2 \rmt B}$ are the degree
                  of coherence and an exchange function due to the
                  Bose-Einstein effect. $f(q)$ is attributed to the
                  resonances
                  effect; $f(q) \approx 0.9\sim 0.7$ depends on the
                  Monte Carlo programs used \cite{wolf94}.
\end{references}
\end{document}